\documentclass[iop]{emulateapj}

\usepackage{natbib} 
\newcommand{\unit}[1]{\mathrm{#1}}

\newcommand{\wpp}{w_{\mathrm{p}}}
\newcommand{\wprp}{w_{\mathrm{p}}(r_{\mathrm{p}})}
\newcommand{\rp}{(r_{\mathrm{p}})}

\newcommand{\Mpc}{\,\unit{Mpc}} 

\newcommand{\Gyr}{\unit{Gyr}}

\newcommand{\kms}{\,\unit{km \ s^{-1}}}

\newcommand{\hmpcvol}{h^{-3}\mathrm{Mpc^{3}}}
\newcommand{\hkpc}{h^{-1}\,\mathrm{kpc}}
\newcommand{\hMsun}{h^{-1}\,M_{\odot}}

\newcommand{\Mstar}{M^{\ast}}
\newcommand{\Mstarzero}{M^{\ast}_{0}}
\newcommand{\Lstar}{L^{\ast}}

\newcommand{\Mone}{M_\mathrm{1}}

\newcommand{\expon}{\mathrm{exp}}

\newcommand{\Msun}{\,M_{\odot}}

\newcommand{\fsat}{f_\mathrm{sat}}

\newcommand{\pimax}{\pi_\mathrm{max}}

\newcommand{\vmax}{\mathrm{v}_\mathrm{max}}
\newcommand{\vmaxacc}{\mathrm{v}_\mathrm{max}^\mathrm{acc}}

\newcommand{\vpeak}{\mathrm{v}_\mathrm{peak}}

\newcommand{\Mpeak}{\mathrm{M}_\mathrm{peak}}
\newcommand{\Macc}{\mathrm{M}_\mathrm{acc}}

\newcommand{\wtheta}{w(\theta)}

\usepackage{epsfig}
\usepackage{graphicx}
\usepackage{rotating}

\begin{document}

\title{the strikingly similar relation between satellite and central
galaxies and their dark matter halos since $\lowercase{z}=2$}

\author{Douglas~F.~Watson\altaffilmark{1,2} and
  Charlie Conroy\altaffilmark{3}}

\altaffiltext{1}{NSF Astronomy \& Astrophysics Postdoctoral Fellow,
  Department of Astronomy \& Astrophysics, The University of Chicago,
  Chicago, IL 60637, USA}
\altaffiltext{2}{Kavli Institute for Cosmological Physics, 5640 South
  Ellis Avenue, The University of Chicago, Chicago, IL 60637, USA}
\altaffiltext{3}{Department of Astronomy \& Astrophysics, University
  of California, Santa Cruz, CA, 95064, USA}



\begin{abstract}

  Satellite galaxies in rich clusters are subject to numerous physical
  processes that can significantly influence their evolution.
  However, the typical $L^\ast$ satellite galaxy resides in much lower
  mass galaxy groups, where the processes capable of altering their
  evolution are generally weaker and have had less time to operate.
  To investigate the extent to which satellite and central galaxy
  evolution differs, we separately model the stellar mass - halo mass
  ($M_\ast-M_h$) relation for these two populations over the redshift
  interval $0<z<1$.  This relation for central galaxies is constrained
  by the galaxy stellar mass function while the relation for satellite
  galaxies is constrained against recent measurements of the galaxy
  two-point correlation function (2PCF).  Our approach does not rely
  on the abundance matching technique but instead adopts a flexible
  functional form for the relation between satellite galaxy stellar
  mass and subhalo mass, where subhalo mass is considered at the
  maximum mass that a subhalo has ever reached in its merger history,
  $\Mpeak$.  At $z\sim0$ the satellites, on average, have $\sim 10\%$
  larger stellar masses at fixed $\Mpeak$ compared to central galaxies
  of the same halo mass (although the two relations are consistent at
  $2-3\sigma$ for $\Mpeak$ $\gtrsim 10^{13}\Msun$).  This is required
  in order to reproduce the observed stellar mass-dependent 2PCF and
  satellite fractions.  At low masses our model slightly
  under-predicts the correlation function at $\sim1$ Mpc scales.  At
  $z\sim1$ the satellite and central galaxy $M_\ast-M_h$ relations are
  consistent within the errors, and the model provides an excellent
  fit to the clustering data.  At present, the errors on the
  clustering data at $z\sim 2$ are too large to constrain the
  satellite model.  A simple model in which
  satellite and central galaxies share the same $M_\ast-M_h$ relation
  is able to reproduce the extant $z\sim 2$ clustering data.  We
  speculate that the striking similarity between the satellite and
  central galaxy $M_\ast-M_h$ relations since $z\sim2$ arises because
  the central galaxy relation evolves very weakly with time and
  because the stellar mass of the typical satellite galaxy has not
  changed significantly since it was accreted.  The reason for this
  last point is not yet entirely clear, but it is likely related to
  the fact that the typical $\sim L^\ast$ satellite galaxy resides in
  a poor group where transformation processes are weak and lifetimes
  are short.

\end{abstract}


\keywords{cosmology: theory --- dark matter --- galaxies: halos ---
  galaxies: evolution --- galaxies: formation --- large-scale
  structure of universe}


\section{INTRODUCTION}
\label{sec:intro}


Understanding the evolution of stellar mass in galaxies is a crucial
piece in the galaxy formation puzzle.  Substantial effort has been
devoted to constraining how the stellar content of galaxies is related
to their dark matter halos, which provides a fundamental bridge
between observation and theory.  Such stellar mass - halo mass
relations ($M_\ast-M_h$) have yielded a variety of novel insights,
including constraints on the efficiency of galaxy formation and
connections between galaxy growth and halo growth over much of cosmic
time \citep{white_zheng07, zheng07, conroy_wechsler09, yang12,
  moster13, behroozi12b, behroozi12, leitner12, wang_etal12}.  There
is a growing consensus that the $M_\ast-M_h$ relation evolves little
over most of cosmic history; the implications of this non-evolution are
only beginning to be explored.

Galaxies can be classified as `central' or `satellite' depending on
whether they are located near the center of a parent dark matter halo
or whether they reside within a larger system.  Satellite galaxies can
lead tumultuous lives as they orbit within the intense gravitational
field of their host halo \citep[e.g.,][]{kravtsov04b}.  A number of
complex processes can therefore affect the stellar mass evolution of
satellite galaxies, including: (1) cold gas stripping due to ram
pressure \citep{gunn_gott72}, (2) `strangulation' \citep{larson80},
which removes the hot gas reservoir surrounding satellites, thereby
reducing the fuel available to feed the cold gas disk on long
timescales, (3) shredding of satellite galaxies due to tidal stripping
\citep{purcell_etal07, watson_etal12b}, and (4) gravitational
interactions with other nearby galaxies known as `harassment'
\citep{moore_etal98}.  These effects leave observational signatures on
features such as galaxy star formation histories, colors,
morphologies, and the amount and spatial distribution of intrahalo
light.  While much attention has focused on the evolution of
satellites in clusters, where these effects tend to be maximally
important, the typical $\sim\Lstar$ satellite galaxy resides in poor
groups \citep{zehavi05a, vdBosch07, zehavi11}.  It may therefore be
the case that the evolution of the typical satellite galaxy is not
markedly different from the average central galaxy, at the same
stellar mass.  Understanding the relation between satellite galaxies
and their dark matter subhalos (self-bound, dark matter structures
orbiting in the potential of their host halo), and how this relation
differs from the relation between central galaxies and their parent
halos, has the potential to constrain the importance of these various
complex physical processes, and is the primary goal of the present
work.

There are two general approaches to constraining the $M_\ast-M_h$
relation.  First, there is the halo occupation distribution (HOD)
technique (and the closely related conditional luminosity function
technique \citealt{yang03}), which adopts flexible functional forms for mapping
galaxies into dark matter halos \citep[e.g.,][]{peacock00a, seljak00,
  scoccimarro01a, berlind02, cooray02, zheng07, leauthaud11a}.
Conceptually, galaxies are separated into centrals and satellites,
with the centrals residing at rest at the center of a parent dark
matter halo.  Satellites are then modeled as orbiting within parent
dark matter halos, with some additional assumptions regarding their
radial and velocity profiles.  This approach does not explicitly use
any information about dark matter subhalos (i.e., satellites are not
constrained to reside within subhalos).  The second technique is known
as abundance matching \citep{kravtsov04a, vale_ostriker04,
  tasitsiomi_etal04, vale_ostriker06, conroy06, conroy_wechsler09,
  guo10, simha10, neistein11a, watson_etal12b, reddick12,
  rod_puebla12, hearin_etal12b}.  In this approach galaxies are
assigned to dark matter halos such that galaxy and halo mass are
monotonically related with the most massive galaxies residing in the
most massive halos.  In this case subhalos are used to place satellite
galaxies in the cosmic web.  The abundance matching technique requires
as an input the spatial abundance of objects, i.e., the luminosity or
mass function of galaxies.  Despite its simplicity, this technique has
been shown to reproduce a variety of observations including galaxy
two-point correlation functions (2PCFs) \citep{conroy06,
  reddick12}, close pair counts \citep{berrier06, berrier_cooke12},
$M_\ast-M_h$ relations \citep{conroy_wechsler09, wang_jing10,
  guo10,reddick12}, and group multiplicity functions
\citep{hearin_etal12b}.  However, several studies have noted
discrepancies between abundance matching predictions and galaxy
statistics at low-$z$ \citep[e.g.,][]{hearin_etal12b} and intermediate
redshifts \citep[e.g.,][]{wetzel_white10,gerke12}, and we expound on
these shortcomings in more detail throughout the paper.  Recently,
\citet[][hereafter B12]{behroozi12} constrained $M_\ast-M_h$ relations
over most of cosmic history using a different technique, where
satellites were placed in dark matter subhalos, and in which the
$M_\ast-M_h$ relation was flexibly parameterized and fit to a suite of
data rather than being derived directly via the abundance matching
formalism (a technique first introduced by \citealt{moster10}).

A key assumption made in most modeling efforts that place satellites
in dark matter subhalos is that the satellite galaxy -- subhalo mass
relation is identical to the central galaxy -- halo mass relation
(with the appropriate choice for the subhalo mass; see Section 2).  In
the present work we relax this assumption and attempt to separately
constrain the satellite $M_\ast-M_h$ relation using galaxy clustering
measurements and, where available, directly estimated satellite
fractions.  This approach has also be taken recently by
\citet{wang06}, \citet{neistein11b}, and \citet{rod_puebla12}. Such an
approach can help address a variety of questions about satellite
galaxies.  For instance, \emph{How well does the stellar mass- halo
  mass relation of satellite galaxies trace the central galaxy
  relation}?  \emph{How does the satellite relation evolve with
  redshift}?  \emph{Is there strong evolution in the peak and
  characteristic mass scale for satellite galaxy star formation
  efficiency}?

The rest of this paper is organized as follows.  In \S~\ref{sec:sims}
we discuss the simulation and halo catalogs used in this study.  In
\S\ref{sec:modeling} we lay out our theoretical framework, in
\S~\ref{sec:results} we present our results, and a discussion is
presented in \S~\ref{sec:discussion}.  Finally, in
\S~\ref{sec:conclusion} we give a summary of our work.  Throughout we
assume a \citet{chabrier03} IMF and a flat $\Lambda$CDM cosmological
model with $\Omega_{\mathrm{m}}=0.27$ and $\Omega_{\Lambda}=0.73$.  Regarding the Hubble constant, we assume $H_0=70$ km s$^{-1}$
Mpc$^{-1}$ throughout, except when comparing to $z\sim0$ clustering
and satellite fraction data where we assume the convention of those
data: $H_0=100\,h$ km s$^{-1}$ Mpc$^{-1}$.


\begin{figure*}
\begin{center}
\includegraphics[width=1.\textwidth]{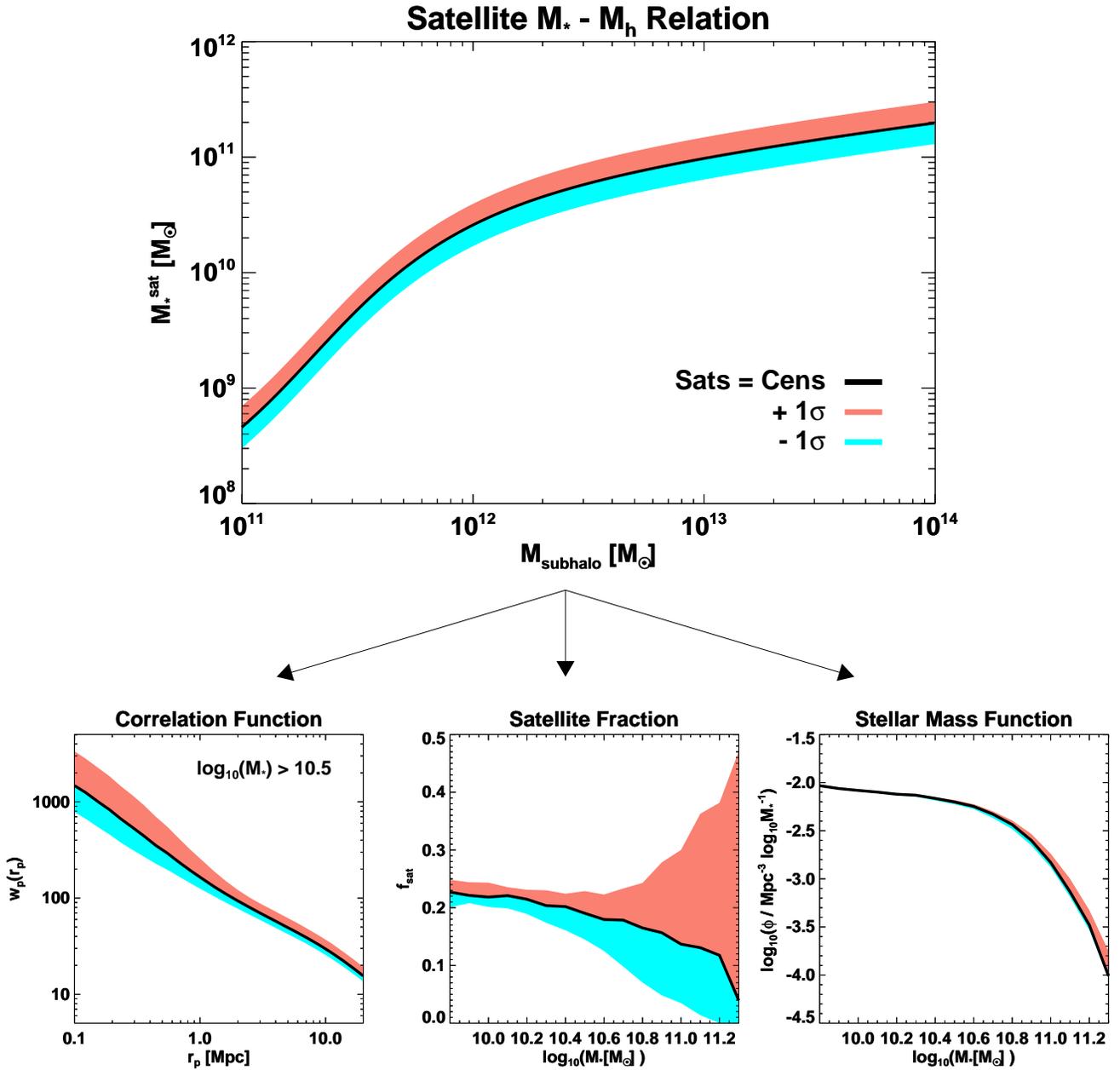}
\caption{``Proof of concept'' demonstrating the effect of varying the
  satellite galaxy $M_\ast-M_h$ relation on the two-point correlation
  function, satellite fraction, and stellar mass function.  The top
  panel shows the satellite stellar mass - halo mass relation at
  $z\sim 0$, where the solid black curve is the \emph{Sats=Cens} model
  of \citet{behroozi12}, such that the satellite $M_\ast-M_h$ relation
  is fixed to the central galaxy relation.  The solid red and blue
  bands show the $1\sigma$ deviation associated with this relation
  ($\sim0.2$ dex scatter).  By populating a high-resolution N-body
  simulation with stellar masses based on this mapping, we compute the
  projected galaxy two-point correlation function for a mass threshold
  log($\Mstar) > 10.5\Msun$ (bottom left panel), the satellite
  fraction (bottom middle panel), and the galaxy stellar mass function
  (bottom right panel).  Both the satellite fraction and the
  correlation function are very sensitive to variation in the
  satellite $M_\ast-M_h$ relation.  In contrast, the stellar mass
  function is hardly affected by the $\pm 1\sigma$ shift in the
  satellite $M_\ast-M_h$ relation because satellite galaxies are a
  minority population.  This allows us to alter the satellite relation
  in order to match clustering measurements and satellite fractions
  (where available) while still ensuring that the model reproduces the
  global stellar mass function.}
\label{fig:motivation}
\end{center}
\end{figure*}



\section{Simulation and Halo Catalogs}
\label{sec:sims}



In the present work we employ the Bolshoi high resolution
collisionless $N-$body simulation of cold dark matter
\citep{bolshoi_11}.  The simulation has a volume of $250\,\hmpcvol$
with $2048^{3}$ particles and a standard cold dark matter
($\Lambda$CDM) cosmological model with $\Omega_{\mathrm{m}}=0.27$,
$\Omega_{\Lambda}=0.73$, $\Omega_{\mathrm{b}}=0.042$, $h=0.7$,
$\sigma_{8}=0.82$, and $n_{\mathrm{s}}=0.95$.  The simulation has a
mass resolution of $1.9\times 10^8\,\Msun$ and force resolution of
$1\,\hkpc$, and particles were tracked from $z=80$ to $z=0$.  Bolshoi
was run with the Adaptive Refinement Tree Code (ART;
\citealt{kravtsovART97, kravtsovklypin99}).

Halos and subhalos were identified with the phase-space temporal halo
finder ROCKSTAR \citep{rockstar,rockstar_trees}.  ROCKSTAR is capable of resolving the maximum circular velocity, $\vmax \equiv \sqrt{GM(<r)/r}|_{\mathrm{max}}$, of halos and subhalos down to a minimum value of $\sim55\kms$.  We use the ROCKSTAR halo catalogs at $z=0.05, 0.9$ and
$1.9$ to model the satellite galaxy $M_\ast-M_h$ relation for
clustering measurements at these same (median) redshifts.  Halo masses
were calculated using spherical overdensities according to the
redshift-dependent virial overdensity formalism of
\citet{bryan_norman98}.  Halo centers are defined as being located at the peak in the phase-space density distribution (see \S~3.5.1 of \citealt{rockstar_trees} for details).

Throughout this work we consider the halo mass definition $\Mpeak$ for
subhalos, which is the maximum mass that a subhalo has ever reached in
its merger history.  As we discuss in \S~\ref{sec:modeling}, our
theoretical framework is built off the $M_\ast-M_h$ relation of B12
which employs $\Mpeak$, thus we adopt this definition for consistency.
$\Mpeak$ is closely related to the commonly used $\Macc$, the mass of
a subhalo at the epoch of accretion\footnote{We have repeated our
  analysis using subhalos at $\Macc$ and find our results to be
  consistent with those based on $\Mpeak$.}.  These quantities are
important because subhalos lose mass due to tidal stripping throughout
their orbital history.  However, the galaxy is typically much more
centrally concentrated than the halo, and so it is expected to undergo
tidal stripping only after its halo has been nearly destroyed.  Thus
we expect the galaxy content of a subhalo to be more closely tied to
the mass of the halo at the epoch of accretion or its maximum mass
throughout its history, rather than the present mass of the halo
\citep[see e.g.,][for further discussion on this
point]{nagai05,vale_ostriker06, conroy06, reddick12}.


\section{THEORETICAL FRAMEWORK}
\label{sec:modeling}



Our approach begins with the $M_\ast-M_h$ relation of B12, which was
constrained to match observational measurements of galaxy stellar mass
functions (GSMFs), star formation rate-stellar mass relations, and the
cosmic star formation history over the redshift range $0<z<8$.  We
retain the B12 relation for central galaxies, since the low-$z$
observed satellite fractions imply that one-point statistics like the
GSMF are primarily sensitive to central galaxies.  In our approach the
satellite $M_\ast-M_h$ relation is constrained by observed 2PCF data
and, where available, direct estimates of the satellite fraction from
group catalogs.

B12 chose a $M_\ast-M_h$ form that was flexible enough to reproduce
the highly constraining low-$z$ measurements, but not so flexible as
to allow for over-fitting of data at higher redshifts where the
uncertainties can be large.  The specific functional form is given as:
\begin{eqnarray}\label{eqn:behroozi}
\mathrm{log}_{10}(M_\ast(M_{h})) & = &\mathrm{log}_{10}(\epsilon \Mone) + f\Big(\mathrm{log}_{10}\Big(\frac{M_\ast}{M_{h}}\Big)\Big) - f(0) \nonumber, \\
\end{eqnarray}
where
\begin{eqnarray}
f(x) & = & -\mathrm{log}_{10}(10^{\alpha x} + 1) + \delta
\frac{(\mathrm{log}_{10\ }(1 + \mathrm{exp}(x)))^{\gamma}}{(1 +
\mathrm{exp}(10^{-x}))}. \nonumber
\end{eqnarray}
The virial mass at the epoch of observation is used for the halo mass
of central galaxies, while $\Mpeak$ is used for subhalos which harbor
the satellites.  The 5 parameters governing the fitting function
are\footnote{Commonly used $M_\ast-M_h$ relations have a double
  power-law form with 4 principal parameters describing a
  characteristic halo/stellar mass and low/high-mass power slopes.
  B12 demonstrated that this double power-law formalism can not
  accurately match GSMFs at various redshifts (see Appendix D of
  B12).}:(1) the characteristic halo mass $\Mone$, (2) the ratio of
the characteristic stellar and halo masses $\epsilon = \Mstarzero /
\Mone$, (3) the low-mass slope $\alpha$, (4) the index of the subpower
law\footnote{A subpower law is a function that is asymptotically
  shallower than any power law, yet asymptotically greater than a
  logarithm.} of the $M_\ast-M_h$ relation high-mass slope $\gamma$,
and (5) the strength of the subpowerlaw at the massive end $\delta$.
There is also scatter in the $M_\ast-M_h$ relation, $\xi$, which is
assumed to be a log-normal distribution and constant with (sub)halo
mass. We will call the model where satellites and centrals are given
the same $M_\ast-M_h$ relation as determined in B12, the
\emph{Sats=Cens} model.  The model in which the satellite $M_\ast-M_h$
relation is allowed to differ from the central relation will be called
the \emph{varySats} model.

The ``proof of concept'' of our technique is shown in Figure~
\ref{fig:motivation}.  The top panel shows the mean satellite galaxy
$M_\ast-M_h$ relation at $z\sim 0$ for the best-fit \emph{Sats=Cens}
model of B12, with the solid red and blue bands depicting the
$1\sigma$ deviation ($\sim0.2$ dex scatter in the $M_\ast-M_h$
relation). We populate the halos and subhalos from the Bolshoi
simulation with stellar masses based on the \emph{Sats=Cens}
prescription and compute the projected 2PCF (bottom left panel), the
satellite fraction (bottom middle panel), and the GSMF (bottom right
panel).  The details of the modeling are discussed below.    The crucial
point is how the $1\sigma$ error propagates into these three statistics.  The
2PCF and the satellite fraction ($\fsat = n_{\rm sat}/n_{\rm tot}$)
are very sensitive to the satellite $M_\ast-M_h$ relation.  For the
2PCF this is especially notable on scales $\mathrm{r}_\mathrm{p}\lesssim
1\Mpc$ (we are showing a mass threshold log($\Mstar) > 10.5\Msun$ as
this corresponds to a mass cut used by \citet{reddick12} that we
compare to throughout the paper). One can think of the 2PCF as a sum
of two terms: on small scales, pairs of galaxies reside in the same
host dark matter halo (the `1-halo' term), whereas on large scales,
the individual galaxies of a pair reside in distinct halos (the
`2-halo' term).  The shape and amplitude of the 2PCF on small scales
($r \lesssim 1 \Mpc$) are thus very sensitive to central-satellite and
satellite-satellite pairs.  $\fsat$ is sensitive to changes in the
satellite $M_\ast-M_h$ relation at all galaxy stellar masses, though
the effect is most marked at the high-mass end.  However, notice that
the GSMF is hardly affected at all by the $\pm 1\sigma$ shift in the
satellite $M_\ast-M_h$ relation since satellite galaxies are
subdominant by number.  The fraction of galaxies that are satellites
at low redshift reaches a maximum value of $\fsat \sim 30\%$ for
luminosity-selected samples \citep[e.g.,][]{zheng05, zehavi11} and
decreases with increasing redshift \citep{zheng07, wake_etal11}.
Therefore, we have some freedom to adjust the satellite $M_\ast-M_h$
relation to match clustering measurements while still remaining within
the observational uncertainties of the global GSMF.

We take the following procedure to constrain the satellite
$M_\ast-M_h$ relation:

\begin{enumerate}

\item{Adopt the $M_\ast-M_h$ relation for the central galaxies derived
  in B12. Parametrize the normalization of the central relation via
  the following prefactor\footnote{In principle, one would opt to
    insert pre-factors in front of all of the parameters associated
    with the B12 central $M_\ast-M_h$ relation.  We have found this to
    be too computationally expensive to put into practice.}:
\begin{eqnarray}\label{eqn:vary_params_cen}
\mathrm{log}_{10}(\epsilon_{\mathrm{cen}}) & = & c_{0} \times
\mathrm{log}_{10}(\epsilon) \nonumber \\
\end{eqnarray}
}

\item{Parametrize the satellite relation such that subhalos are
    labeled according to their peak mass, $\Mpeak$, and pre-factors
    have been inserted in front of all the B12 $M_\ast-M_h$ relation
    parameters (as well as the scatter) such that,

\begin{eqnarray}\label{eqn:vary_params}
\mathrm{log}_{10}(M_{1,\mathrm{sat}}) & = & c_{1} \times \mathrm{log}_{10}(\Mone) \nonumber \\ 
\mathrm{log}_{10}(\epsilon_{\mathrm{sat}}) & = & c_{2} \times \mathrm{log}_{10}(\epsilon) \nonumber \\
\alpha_{\mathrm{sat}} & = & c_{3} \times {\alpha} \nonumber \\
\gamma_{\mathrm{sat}} & = & c_{4} \times {\gamma} \nonumber \\
\delta_{\mathrm{sat}} & = & c_{5} \times {\delta} \nonumber \\
\xi_{\mathrm{sat}} & = & c_{6} \times {\xi}. \nonumber \\
\end{eqnarray}
}

\item{Start with the \emph{Sats=Cens} model parameter values (i.e.,
    $c_{i=1...6} = 1.0$) and populate the halos and subhalos of the
    Bolshoi simulation with stellar masses based on this $M_\ast-M_h$
    relation.}

\item{Compute the real-space two-point correlation function from the
    simulation for a given stellar mass threshold sample.  For
    comparison to data, convert either to the projected, $\wprp$, or
    angular correlation function, $\wtheta$.  The former is computed
    via:
\begin{equation}
  \wpp\rp = 2\int_0^{\pimax}\xi\Big(\sqrt{r_{\mathrm{p}}^2+\pi^2}\Big)d\pi,
\end{equation}
with the same upper limit of integration as used in the observational
samples.  The angular correlation function is computed using the Limber
transformation,
\begin{equation}
  \wtheta =
  \frac{2}{c}\int_{0}^{\infty}dzH(z)N(z)^{2}\int_{0}^{\infty}\xi(
  \sqrt{u^{2} + D_{m}(\bar{z})^{2}\theta^{2}})du,
\end{equation}
where \emph{c} is the speed of light, $H(z)$ is the Hubble constant at
redshift $z$, $N(z)$ is the normalized redshift distribution of the
galaxies in the sample, and $D_{m}(\bar{z})$ is the comoving distance
to the median redshift.}

\item{Compute the likelihood between the observed
    and predicted clustering for all available stellar mass threshold
    samples.  Do this separately, not simultaneously, for the three redshifts considered: $z\sim0, 1$ and 2.  Incorporate the measured satellite fraction as an
    additional constraint where available\footnote{Direct estimates of
      satellite fractions are not available at $z\sim 1$ and $2$ and
      so this additional constraint is not considered at those epochs.
      It is also true that $\wprp$ and $\fsat$ are highly covariant.
      We do not take this into account in our modeling, though this
      should be incorporated in future work.}.}

\item{Use a Markov Chain Monte Carlo (MCMC) method, varying the
  parameters $c_{i=0-6}$, to compute the range of possible satellite
  $M_\ast-M_h$ relations given the data. Specifically, adopt the
  Metropolis-Hastings algorithm, which works as follows.  With the
  initial free parameter values from step~3, compute $\chi^2$ from
  this starting point.  Steps are then chosen for the parameters and
  $\chi^2$ is computed for the new location.  This new location is
  added to the chain if
  $\chi^{2}_{\mathrm{new}}<\chi^{2}_{\mathrm{old}}$ or $n <
  \expon[-(\chi^{2}_{\mathrm{new}}-\chi^{2}_{\mathrm{old}})/2]$, where
  n is a random number between 0 and 1.  If these conditions are not
  met, then the old location is repeated in the chain.  This process
  then repeats until the chain has converged\footnote{We incorporate
    the full covariance matrices for our model fitting at $z\sim 0$
    and 2.  Covariance matrices are not available for the DEEP2
    $z\sim1$ samples.  We assume non-informative, uniform priors
    on the parameters and we start numerous chains at a wide range of
    initial $c$ pre-factor values to confirm that they converge to the
    same parameter values.  For more
    details on MCMC techniques, see \citet{dunkley05}.}.}

\end{enumerate}

In the present work we focus on the satellite $M_\ast-M_h$ relation at
$z\sim0$, 1, and 2.  At $z\sim0$ we consider a recent measurement of
the stellar mass-dependent 2PCF derived by \citet{reddick12} using the
New York University Value Added Galaxy Catalog \citep{VAGC_05}, based
on the Sloan Digital Sky Survey (SDSS) Data Release 7
\citep{padmanabhan08,DR7_09}.  We also use satellite fractions derived
in Reddick et al. from SDSS group catalogs.  The group catalogs were
constructed according to the methodology described in
\citet{tinker13}.  At $z\sim1$ we employ two-point clustering
measurements from the DEEP2 Galaxy Redshift Survey
\citep{DEEP2newman12} by \citet[][hereafter M12]{mostek12}.  At
$z\sim2$ constraints are provided by angular clustering measurements
from the NEWFIRM Medium Band Survey (NMBS;
\citealt{NEWFIRMvonDokkum09}) as derived in \citet{wake_etal11}.  In
the following section we confront our model with these data.


\begin{figure*}
\begin{center}
\includegraphics[width=1.\textwidth]{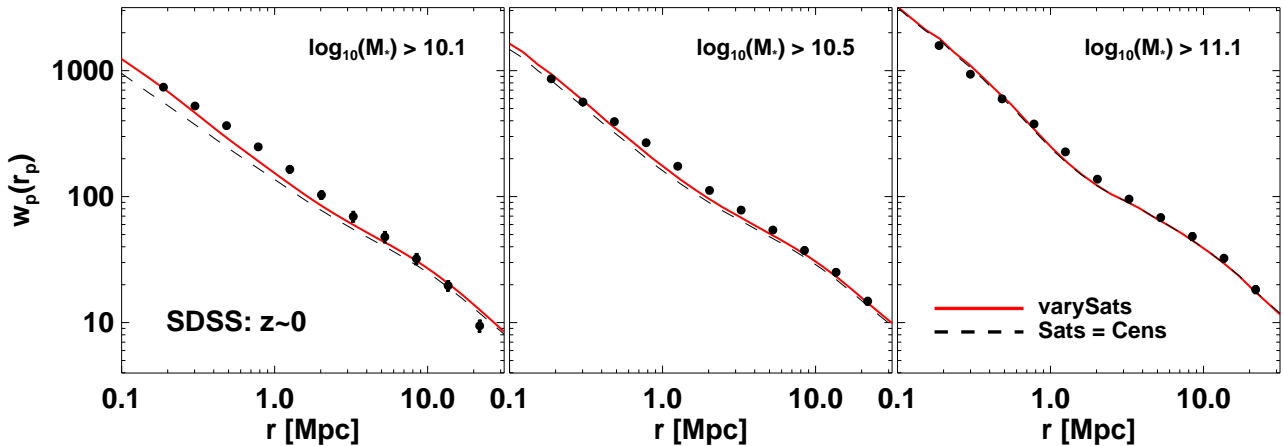}
\caption{Comparison between model and observed projected correlation
  function as a function of stellar mass for three SDSS threshold
  samples as measured by \citet{reddick12}: log$(\Mstar) > 10.1, 10.5,
  11.1 \Msun$.  The black dashed curve shows the \emph{Sats=Cens}
  model prediction where the satellite $M_\ast-M_h$ relation is fixed
  to that of the central galaxies. The red solid curve shows our
  best-fit model in which the $M_\ast-M_h$ relation for satellite
  galaxies is allowed to vary to better match observed clustering and
  satellite fractions (see Figure~\ref{fig:smf_fsat}).  There is still
  tension between the \emph{varySats} model and the data at the 1-2
  halo transition, especially at lower masses.}
\label{fig:wp_reddick}
\end{center}
\end{figure*}



\section{Results}
\label{sec:results}


We now compare the \emph{Sats=Cens} and \emph{varySats} models to
observational constraints at $z\sim0$ and $1$, and the
\emph{Sats=Cens} model to data at $z\sim2$.  We derive satellite
$M_\ast-M_h$ relations at $0<z<1$ and investigate its evolution and
how it compares to that of central galaxies. We also present satellite
fractions over the interval $0<z<2$.


\subsection{Clustering and the Satellite $M_\ast-M_h$ Relation at
  $\lowercase{z}\sim0$}
\label{subsec:mod_z0}



\begin{figure}
\begin{center}
\includegraphics[width=.5\textwidth]{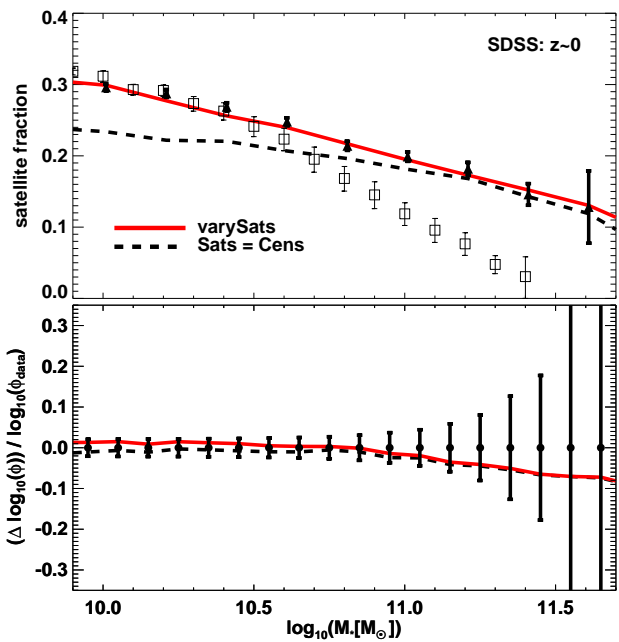}
\caption{ \emph{Top panel}: Best-fit model satellite fraction (red
  curve) as a function of galaxy stellar mass (black data points with
  error bars are from \citealt{reddick12}).  The \emph{Sats=Cens}
  model clearly fails to reproduce the measured satellite fractions at
  low masses, while the \emph{varySats} model is in good agreement
  with the data.  We also show the satellite fractions as measured by \citet{yang09a} as open black boxes and in the text we discuss the implications of the discrepancy at log$(M_\ast) \gtrsim 10.7$ on our modeling results.  \emph{Bottom panel}: Residuals between the observed
  and model galaxy stellar mass function; data are from
  \citet{baldry08}.  The galaxy stellar mass function is dominated by
  central galaxies, so varying the satellite galaxy $M_\ast-M_h$
  relation has little effect on the galaxy stellar mass function and
  is within the errors.}
\label{fig:smf_fsat}
\end{center}
\end{figure}



\begin{figure}
\begin{center}
\includegraphics[width=.5\textwidth]{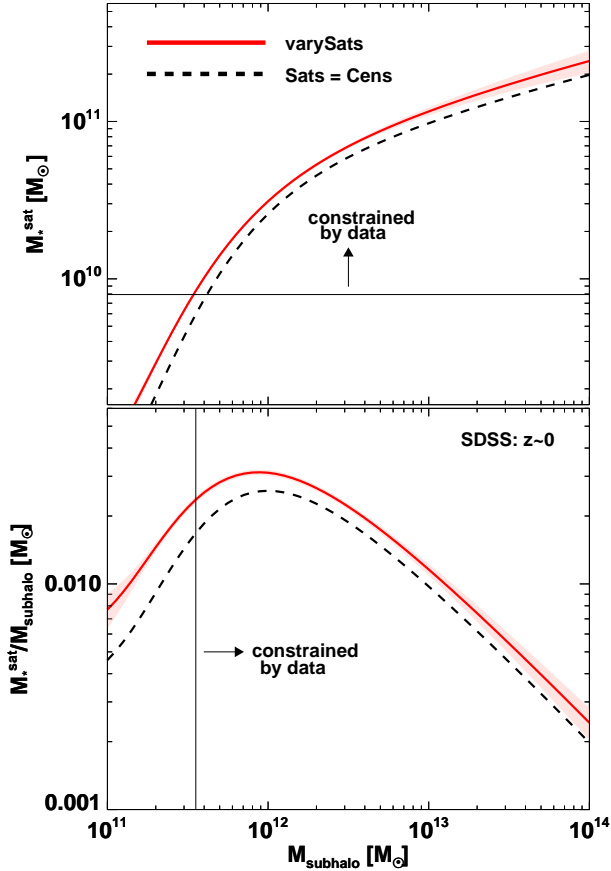}
\caption{The satellite galaxy $M_\ast-M_h$ relation at
  $z\sim0$. \emph{Top panel}: The black curve represents the
  \emph{Sats=Cens} model.  The thin solid lines demarcate the mass
  range over which the model has been constrained by data.  The red
  curve shows the best-fit relation for the \emph{varySats} model.
  The lighter red band represents the $1\sigma$ uncertainty in the
  model. The \emph{varySats} model is driven to higher masses in order
  to boost the clustering and satellite fractions.  \emph{Bottom
  panel}: Ratio between stellar mass and halo mass as a function of
  halo mass (the integrated star formation efficiency). The peak in
  star formation efficiency and the characteristic (sub)halo mass is
  nearly identical in both cases highlighting the mild stellar mass
  evolution in satellite galaxies and their similarity to the known
  lack of evolution of central galaxies.}
\label{fig:sat_shmr_z0}
\end{center}
\end{figure}



\begin{figure*}
\begin{center}
\includegraphics[width=1.\textwidth]{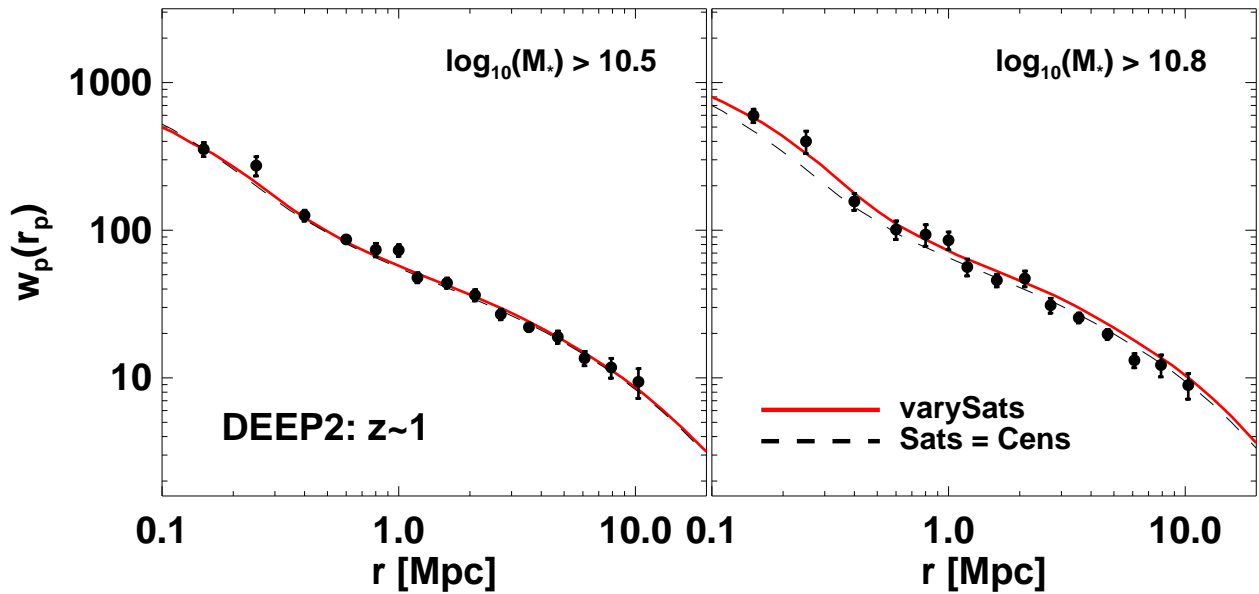}
\caption{Same as Figure~\ref{fig:wp_reddick} but for $z\sim 1$ DEEP2
  measurements of the projected correlation function of galaxies as a
  function of stellar mass from \citet{mostek12}.  Both models provide
  an excellent fit to the data.}
\label{fig:wp_mostek}
\end{center}
\end{figure*}



\begin{figure}
\begin{center}
\includegraphics[width=.5\textwidth]{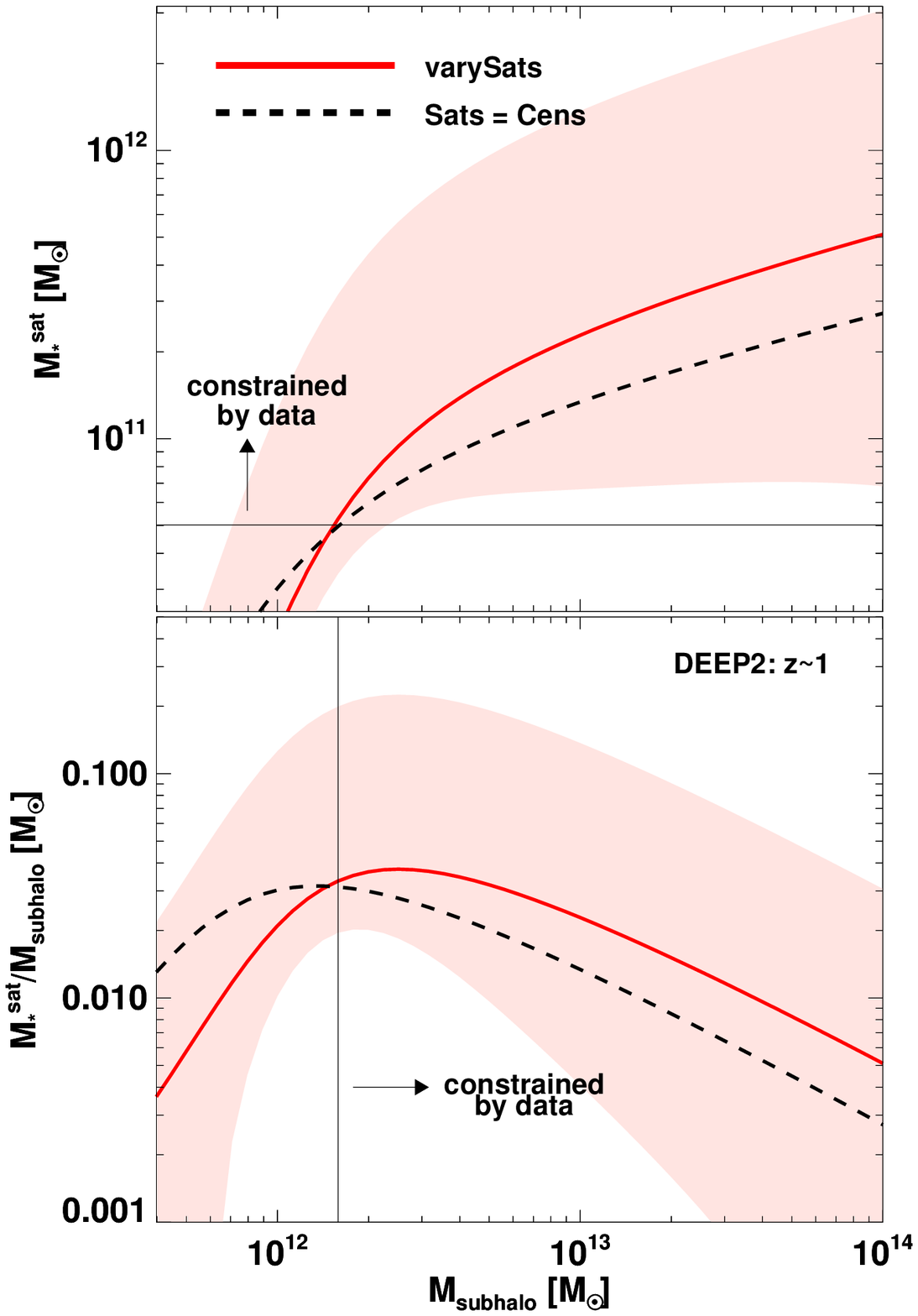}
\caption{Same as Figure~\ref{fig:sat_shmr_z0} but for $z\sim 1$. The
  errors on the $z\sim1$ \emph{varySats} model are considerably larger than at
  $z\sim0$, with the result that the $M_\ast-M_h$ relation for this
  model is consistent with the relation for the simple
  \emph{Sats=Cens} model.}
\label{fig:sat_shmr_z1}
\end{center}
\end{figure}


Figure~\ref{fig:wp_reddick} shows the projected 2PCF at $z\sim0$ at
three separate stellar mass thresholds: log$(\Mstar) >$ 10.1, 10.5, and
11.1 $\Msun$.  The black dashed curves show the \emph{Sats=Cens}
model where the clustering is computed from the best-fit $M_\ast-M_h$
relation given by B12.  We emphasize that this curve is \emph{not} a
fit to the clustering data, rather it is a direct prediction of the
B12 $M_\ast-M_h$ relation.  The \emph{Sats=Cens} model provides an
excellent fit to the data at the largest stellar mass thresholds and
on large scales, but under-predicts the clustering on small scales in
the two lower mass samples.

This under-prediction is probably not due to resolution effects in the
Bolshoi simulation, as discussed in Appendix B of \citet{reddick12}.
In addition, \citet{watson_etal12b} used the \citet{zentner05}
semi-analytic model for subhalos, which has the capability of tracking
subhalos down to circular velocities $\sim 0 \,\kms$, to test
resolution effects in numerical simulations.  They performed a test in
which they selected a $\vmaxacc > 210\,\kms$ threshold (corresponding
to $\sim \Lstar$ galaxies and brighter) with two additional $\vmax$
thresholds of $\vmax > 20\,\kms$ and $\vmax > 80 \,\kms$ to mimic the
effect of a resolution limit of a given simulation.  Increasing this
threshold from $20$ to $80\,\kms$ resulted in a maximum $\sim 20\%$
decrease in the 2PCF at scales $r \lesssim 1 \Mpc$.  Increasing from
$20$ to $40 \kms$ only had a few percent effect on the small-scale
amplitude.  Though not explicitly shown in their paper, they repeated
this test for $\vmaxacc > 170\,\kms$ and found nearly the same result,
implying that the resolution limit of $55 \kms$ should not have a
significant effect on the 2PCF for the lowest stellar mass threshold
considered in Figure \ref{fig:wp_reddick}.

We have also constructed 1000 \emph{Sats=Cens} models drawing from the
errors in the B12 best-fit parameters and found the $1\sigma$
distribution in the 1000 2PCFs to still be significantly discrepant
from the data on small scales at lower masses.  Thus, no
\emph{Sats=Cens} model can provide a satisfactory fit to the
small-scale clustering for the lowest stellar mass threshold sample.

The \emph{Sats=Cens} model succeeds on large scales because the
large-scale clustering is only weakly sensitive to the treatment of
satellite galaxies.  On smaller scales the 2PCF is much more sensitive
to the relation between satellite galaxy mass and halo mass.  The
\emph{varySats} model allows for freedom in the satellite $M_\ast-M_h$
relation and thus is able to more closely match the data shown in
Figure \ref{fig:wp_reddick}, though there is clearly still unresolved
tension at the 1- and 2-halo junction at $\sim1 \Mpc$.

In addition to the 2PCF, the \emph{varySats} model is also constrained
by the measured satellite fraction, shown in Figure
\ref{fig:smf_fsat}.  As we see in the top panel of Figure
\ref{fig:smf_fsat}, the data require a larger satellite fraction for
log$(\Mstar) < 10.9\Msun$ than the prediction of the \emph{Sats=Cens}
model.  The open squares are the measurements of \citet{yang09a} for
comparison to the R12 results.  We use R12 throughout the paper for
consistency with the simulation and data used in B12.  However,
group-finder effects can lead to large biases at large $M_\ast$ and
the \citet{yang09a} data are clearly discrepant with R12 at
log$(M_\ast) \gtrsim 10.7$.   Specifically, this arises because R12
does not assume that central galaxies are necessarily the most massive
(or brightest) in their host halo (see \S~7.3 of R12 for details).
The disagreement becomes relevant at very large stellar masses,
log$(M_\ast) \gtrsim 10.7$.  Thus, this may affect
our modeling of the 2PCF for the largest stellar mass threshold we consider, though our general results should hold.  The bottom panel illustrates that the
changes to the global GSMF of \citet{baldry08} (the same GSMF used in
B12) implied by the \emph{varySats} model are small and within the
error bars.

To evaluate whether (and to what degree) the \emph{varySats} model is
preferred over the \emph{Sats = Cens} model we employ the Aikake
Information Criterion (AIC; see \citealt{liddle06} for details).  AIC
is preferred over a reduced $\chi^{2}$ comparison since we do not take
into account the covariance between $\fsat$ and $\wprp$, which makes
it difficult to discuss the goodness-of-fit in an absolute sense.
Instead, using AIC we are able to comment on model preference. AIC is
defined as AIC$=2k + \chi^2_{\rm min}$, where $k$ is the number of
free parameters in the model and in our case $\chi^{2} =
\chi^{2}(\fsat) + \chi^{2}(\wprp)$.  It is stressed in the literature
to add a correction factor to AIC to take into account small sample
size, such that $\mathrm{AIC_{c}} = \mathrm{AIC} + 2k(k+1)/(N-k-1)$,
where N is the number of data points ($N=44$ for the $z\sim0$
case). For the \emph{Sats = Cens} model, $k=0$, $\chi^{2}_{Sats=Cens}
= 220.1$, the correction factor is 0, so
$\mathrm{AIC_{c,}}_{Sats=Cens}=220.1$. For the \emph{varySats} model,
$k=7$, $\chi^{2}_{varySats} = 64.5$, the correction factor is 3.1,
thus $\mathrm{AIC_{c,}}_{varySats}=81.6$.  When judged on Jeffreys
scale for interpretation \citep{jeffreys61}, the $\Delta \mathrm{AIC_{c}}$ implies that
the \emph{varySats} model preference is considered `decisive'.  Best-fit parameter values and the associated errors from our \emph{varySats} model are listed in Table~\ref{model_table}.


\begin{figure*}
\begin{center}
\includegraphics[width=1.\textwidth]{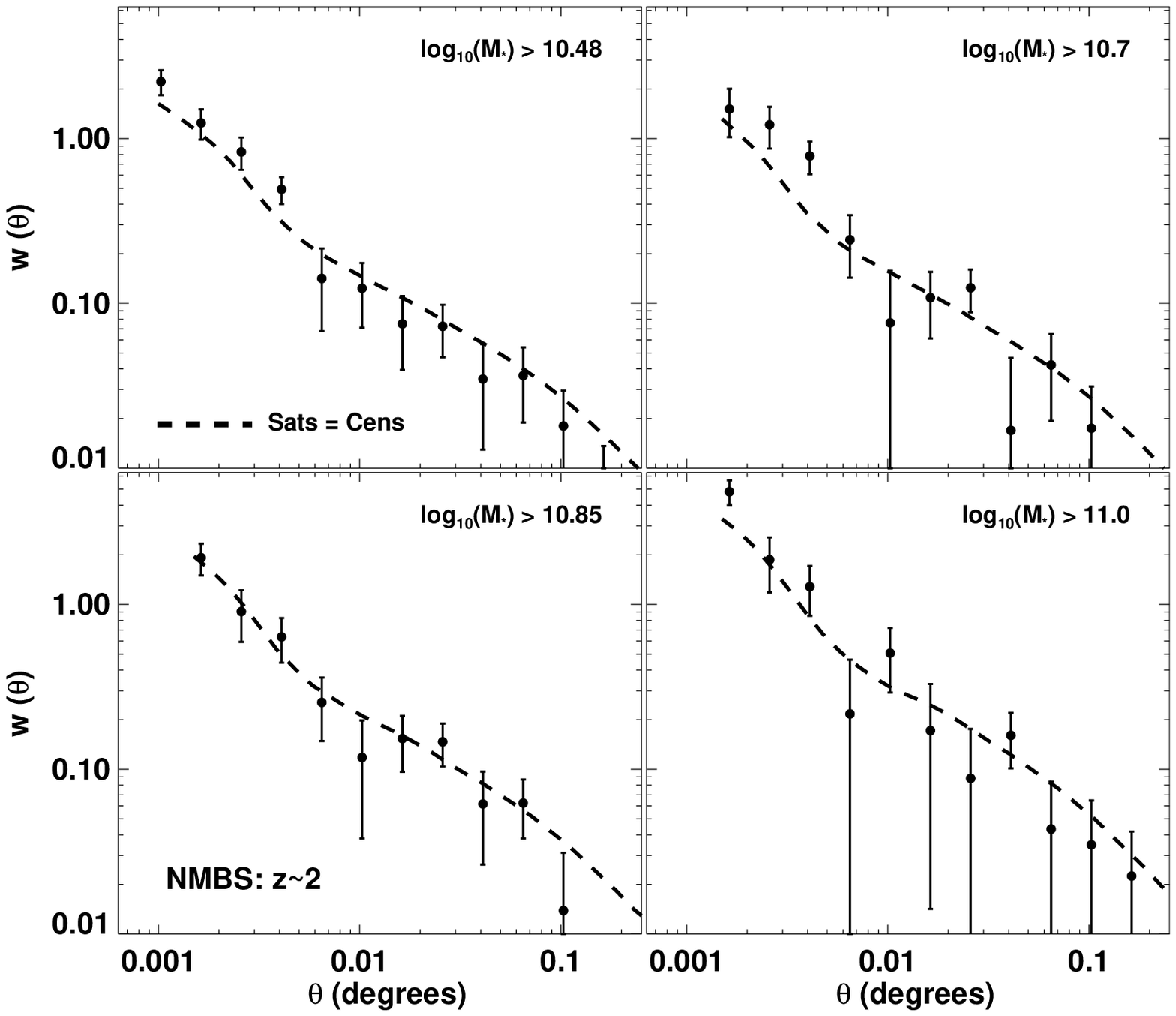}
\caption{Same as Figures \ref{fig:wp_reddick} and \ref{fig:wp_mostek}
  but for $z\sim 2$ NMBS measurements of the angular correlation
  function of galaxies as a function of stellar mass from
  \citet{wake_etal11}.  The \emph{Sats=Cens} model is in remarkably
  good agreement with the data.  Due to the relatively large error
  bars on the data the \emph{varySats} model is unconstrained and so
  is not included in the figure.}
\label{fig:wp_wake}
\end{center}
\end{figure*}


Turning to Figure \ref{fig:sat_shmr_z0}, the top panel shows the
satellite $M_\ast-M_h$ relation: satellite galaxy stellar mass
vs. subhalo \emph{peak} mass, $\Mpeak$.  Again, the dashed black curve
is the \emph{Sats=Cens} model of B12, and the red curve shows the
best-fit result from the \emph{varySats} model.  The solid red band
shows the $1\sigma$ uncertainty in our \emph{varySats} model.  The
\emph{varySats} model implies that above the completeness limit of
log($\Mstar) > 10.1\Msun$ of \citet{reddick12} there should be, on
average, more stellar mass in satellite galaxies than the
\emph{Sats=Cens} model prediction by $\sim 10\%$ at all subhalo masses
where the model can be constrained (although the two relations are
consistent at the $2-3\sigma$ level for subhalo masses $\gtrsim
10^{13}\hMsun$).  This implies that there will be more satellite
galaxies in each SDSS stellar mass $\wprp$ threshold sample, which
will boost the clustering and the satellite fractions.

The bottom panel shows the fraction of available baryons that have
turned into stars (the integrated star formation efficiency) to
illustrate at what subhalo mass galaxy formation is most efficient.
Above the stellar mass limit where the model can be constrained, star
formation in the \emph{varySats} model has been slightly more
efficient, on average, at all subhalo masses than central galaxies of
the same mass.  On the other hand, the characteristic mass for peak
satellite galaxy star formation is approximately the same for central
and satellite galaxies, though we urge caution with this conclusion as
we are unable to probe the full range of the ``turn over" in the
$M_\ast-M_h$ relation.


\subsection{Clustering and the Satellite $M_\ast-M_h$ Relation at $\lowercase{z}\sim1$}
\label{subsec:mod_z1}


As we turn to the $z\sim 1$ results, we find that the \emph{Sats=Cens}
model is in rather good agreement with the $z\sim1$ DEEP2 data as
shown in Figure~\ref{fig:wp_mostek}.  Again, the black dashed curve
shows the \emph{Sats=Cens} model prediction and the red curve is our
best-fit \emph{varySats} model.  Measured satellite fractions
were proved to be a powerful additional constraint at $z\sim0$, however
measurements are not available at $z\sim1$.  We find that $c_0, c_1$, and $c_6$ are reasonably well constrained.  However, $c_2, c_3, c_4$, and $c_5$ are all
unconstrained, thus parameter values (with errors) from our
\emph{varySats} model are not included in Table~\ref{model_table}.

This is lack of constraining power is clearly manifested in Figure \ref{fig:sat_shmr_z1}  where we
show the best-fit $M_\ast-M_h$ relations.  Here the errors on the
satellite relation are significantly larger than at $z\sim0$ (again,
the light red band represents the $1\sigma$ uncertainty in the model),
because the clustering data carry larger errors, only consider very
large $M_*$ thresholds, and because they are no measured satellite
fractions to include in the fitting.  In fact, we have found that the
$c_{0-6}$ parameter values
corresponding to the extrema of the $1\sigma$ error band in
Figure~\ref{fig:sat_shmr_z1} yield $\wprp$ predictions that are in
good agreement with the data.  They also appear to be reasonable when
we test a much lower threshold for $\wprp$ of log($M_\ast) > 9.5$
where data are not available (i.e., there is a relative decrease in
the overall amplitude of $\wprp$ as expected, yet no incongruous
features).  This reinforces the fact that it is not simply $\wprp$
measurements spanning a larger range in $M_\ast$ thresholds that is
necessary to constrain the satellite relation, rather the addition of
measured satellite fractions as an additional, powerful constraint.
Nonetheless, it is clear that the satellite relation is similar to the
central relation at $z\sim1$.  Future data from larger area surveys at
$z\sim1$ should provide stronger constraints on the satellite
$M_\ast-M_h$.


\begin{centering}
\begin{deluxetable*}{c|ccccccc}
\tablecaption{BEST-FIT PARAMETERS FOR THE \emph{varySats} MODEL}
\tablewidth{.9\textwidth} \startdata \hline 	\hline \\[-2ex]
$\mathrm{redshift}$ &  $c_{0}$ &  $c_{1}$  &  $c_{2}$    &   $c_{3}$  & $c_{4}$
&   $c_{5}$  & $c_{6}$ \\ \hline \\[-2ex] $z = 0$  &
$1.024^{+0.011}_{-0.058}$   &   $0.957^{+0.001}_{-0.009}$     &  $0.992^{+0.001}_{-0.002}$     &
$0.732^{+0.156}_{-0.0145}$     &  $1.029^{+0.092}_{-0.036}$     &
$0.986^{+0.067}_{-0.192}$     &  $0.834^{+0.060}_{-0.074}$
 \\ [-2ex] \enddata
\tablecomments{Best-fit parameter values from our \emph{varySats}
  model at $z=0$.  The errors on the parameter values
  represent the extrema of the middle $68.3\%$ of our MCMC chains
  after sorting by $\chi^2$.  The \emph{varySats} model at $z\sim1$ could not be constrained, hence no results are shown at that redshift.  Due to the large errors on the data at $z\sim2$ we do not attempt to constrain the \emph{varySats} model.\\}
\label{model_table}
\end{deluxetable*}
\end{centering}



\subsection{Clustering at $\lowercase{z}\sim2$}
\label{subsec:mod_z2}


\citet{wake_etal11} recently presented stellar-mass dependent angular
correlation functions of galaxies at $z\sim2$ based on the NMBS.  The
errors on the measurements are relatively large (compared to lower
redshift) owing to the relatively small area of the NMBS fields.  For
this reason we did not attempt to fit the \emph{VarySats} model to the
data.  Instead we simply compared the \emph{Sats=Cens} model
prediction to the existing clustering data.  The result is shown in
Figure \ref{fig:wp_wake}.  Again we emphasize that the
\emph{Sats=Cens} model makes a prediction for the clustering of
galaxies, as this model was only tuned to fit the $z\sim2$ GSMF.  The
model fits the data remarkably well over all scales and all mass
ranges probed by the data.  Future surveys covering a larger volume
will be necessary to make more quantitative statements regarding the
similarity of the satellite and central $M_\ast-M_h$ relations.


\begin{figure}
\begin{center}
\includegraphics[width=.5\textwidth]{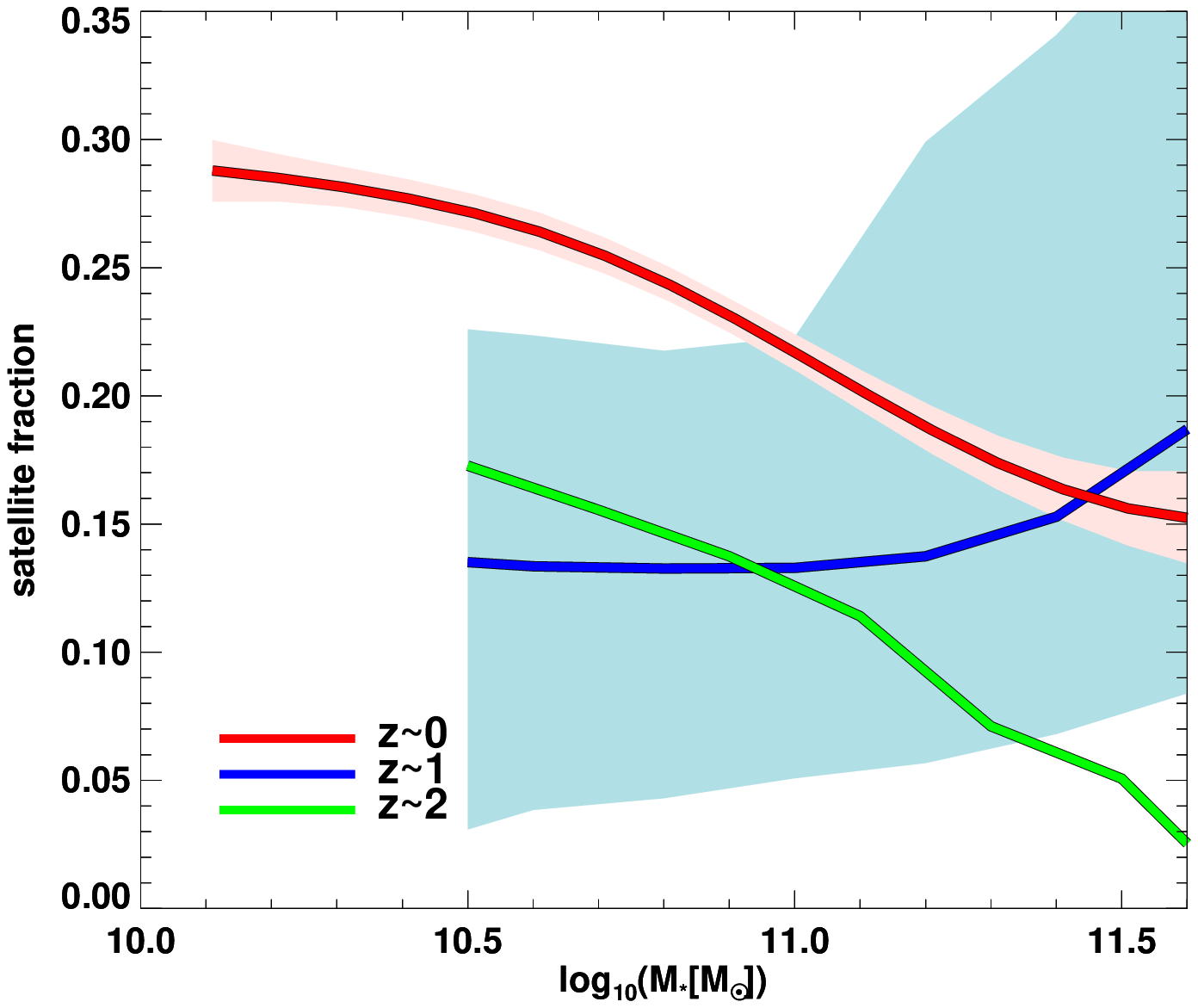}
\caption{Satellite fraction as a function of galaxy stellar mass in
  bins of $\Delta$log($\Mstar) = 0.1$ as inferred from our best-fit
  \emph{varySats} model at $z\sim 0$ and 1 and from the
  \emph{Sats=Cens} model at $z\sim 2$.  Satellite fractions are only
  plotted over the mass range where there is constraining data.}
\label{fig:fsat_z012}
\end{center}
\end{figure}



\subsection{Satellite Fractions at $z\sim 0, 1$ and $2$}
\label{subsec:fsat}


Our modeling describes how galaxy stellar mass should be mapped to
host halos and subhalos.  This mapping can be used to infer satellite
fractions as a function of stellar mass and redshift.
Figure~\ref{fig:fsat_z012} shows the satellite fraction where
constraining data exists as a function of galaxy stellar mass from our
best-fit \emph{varySats} model at $z\sim 0$ and 1 (red and blue
curves).  The green curve is the \emph{Sats=Cens} model at $z\sim2$.
The solid bands are the associated $1\sigma$ errors.  The
\emph{varySats} model was not considered at $z\sim2$, thus no errors
are shown for this curve.  Satellite fractions decline with increasing
redshift in agreement with previous studies \citep{zentner05, zheng07,
  wetzel_white10, watson_powerlaw11}.


\section{DISCUSSION}\label{sec:discussion}



\subsection{Implications for Satellite Galaxy Evolution}
\label{subsec:interp}


The primary result of this paper is that the mean relation between
galaxy stellar mass and halo mass is similar for central and satellite
galaxies since at least $z\sim2$.  We are able to put strong
constraints on the satellite $M_\ast-M_h$ relation at $z\sim0$, and we
find that there is a slight difference in the sense that satellites
appear to have somewhat higher stellar masses at fixed halo mass.  At
$z\sim1$ and $2$ the satellite $M_\ast-M_h$ relation is consistent
with the central relation, but there is still considerable room for
differences between the two relations due to the lack of measured
satellite factions and large errors on the clustering data.  In this section we
explore the implications of these results for the evolution of
satellite galaxies.

There is an emerging consensus that the global $M_\ast-M_h$ relation
evolves little if at all since $z\sim2$ \citep[e.g.,][]{yang12, moster13,
  behroozi12, wang_etal12, leauthaud11a}.  As discussed in previous
sections, the satellite fraction in mass-limited samples is always
$\lesssim30\%$ and decreases with increasing redshift.  The global
relation is therefore, to a good approximation, the central galaxy
relation.  The weakly evolving $M_\ast-M_h$ relation for central
galaxies is one of the main facts that explains the lack of evolution
in the satellite galaxy $M_\ast-M_h$ relation.  In the discussion that
follows we will assume for simplicity that the central relation is
constant in time.

First, it is worth briefly considering scenarios that are inconsistent
with our results.  Satellite galaxies, on average, cannot have
experienced large (order unity) growth in stellar mass since their
accretion, nor can they have experienced a large loss in mass (unless
they are altogether destroyed).  These possibilities would result in
satellite $M_\ast-M_h$ relations significantly offset from the central
galaxy relation.  Of course, since the resulting satellite relation is
an integral over the past history of the satellite, in principle our
results allow for a scenario in which the average satellite has {\it
  both} gained substantial mass via e.g., star formation {\it and}
lost substantial mass via e.g., tidal stripping.  This seems
implausible, at least for the average satellite.  It is therefore
clear that any successful scenario must allow for only modest growth
in the stellar mass of satellite galaxies while they are satellites.

In another scenario for satellite galaxy evolution, stellar mass
growth is immediately halted when a galaxy becomes a satellite.  In
this case the satellite $M_\ast-M_h$ relation would then be identical
to the unevolving central relation.  This would be so because for any
scenario the subhalo peak mass, $\Mpeak$, does not evolve after
accretion by definition, and in this scenario neither does $M^\ast$.
Such a possibility is reminiscent of previous generations of
semi-analytic models.  This is because in such models it was assumed
that, upon infall, a satellite galaxy will be instantaneously stripped
of  its hot gas atmosphere. Consequently, it experiences quenching
after a cold gas consumption time (which typically is very short) and
little $M_\ast$ growth could occur.  However, these earlier models
have been shown to overproduce the fraction of quiescent satellites as
a function of mass \citep[e.g.,][]{weinmann06, Font08}.

We now turn to a more plausible scenario for satellite galaxy
evolution.  In this scenario, all satellites continue to grow in
stellar mass as if they were still central galaxies for a significant
fraction of their lifetime \citep[see also][]{wetzel_etal12b}.  If we
adopt the relation between star formation rate (SFR) and stellar mass
measured in the local universe by \citet{Salim07} along with a typical
time since accretion for satellites of 4 Gyr \citep[see
e.g.,][]{zentner05, wetzel_etal12b}, then the average galaxy with
$M^\ast\sim10^{11}\Msun$ will have increased its mass by $\sim$10\% since
becoming a satellite, as a result of its star formation continuing as if it were a central galaxy.  Owing to the fact that the SFR/$M^\ast$ relation is
so shallow, the fraction of mass grown is a weak function of mass.
Thus, in this scenario, the satellite $M_\ast-M_h$ relation would be
shifted approximately $10\%$ higher compared to the central relation
(i.e., at fixed halo mass satellites would have $\sim10\%$ more
stellar mass).  It is interesting that this shift is similar to what
we find herein at $z\sim0$.  Given the systematic uncertainties that
exist at the $10\%$ level in these relations we caution against
over-interpreting this similarity, although we will now briefly
explore why satellites may evolve like centrals for most of their
evolution.

Satellite galaxies can be subject to numerous physical processes
capable of altering their evolution.  This has been well-documented in
very dense environments, such as large galaxy groups and clusters.
For instance, ram pressure stripping can reduce
the gas supply, stifling star formation. However, these processes are
less effective in modest-sized groups where most satellites reside.
In a given mass-limited sample of galaxies, most satellites are found
in relatively quiescent environments, e.g., in a halo that contains
one central and one satellite galaxy.  This is evident in the group
multiplicity function \citep{berlind06, hearin_etal12b} and in halo
occupation models \citep{zehavi05a, vdBosch07, zehavi11}.  Thus, for
the typical satellite, physical processes capable of altering its
evolution are relatively inefficient.

Perhaps more importantly, dynamical friction timescales at the group
scale are shorter than at the cluster scale (for a satellite of the
same mass), meaning that satellite galaxies in groups spiral toward
the center and merge with the central galaxy more quickly than in rich
clusters.  This implies that there is a relatively small amount of
time for the group environment to shape the physical properties of a
satellite before that satellite is destroyed and/or merges with the
central object.  This effect becomes even stronger at higher redshifts
because dynamical times are shorter (although this effect may be mitigated to some extent due to the fact the SFRs are higher at higher redshift).  This also partly explains the lower
satellite fractions at higher redshifts discussed in Section
\ref{subsec:fsat}.  However, we caution that in this work we are
mainly considering massive satellites.  The situation at lower masses is unconstrained in the present work \citep[see][]{pasquali10}.

This picture was motivated in part by the recent results of
\citet{wetzel_etal12b}, who analyzed the properties of low redshift
satellite and central galaxies including stellar masses and star
formation rates. These authors concluded that, after it is accreted,
the typical satellite continues to form stars as if it were still a
central galaxy for $2-4$ Gyr before star formation is quenched.  This
conclusion follows naturally from the fact that there exists a strong
bimodality in satellite galaxy SFRs combined with the fact that the
locations of the two peaks are similar to those of central galaxies
with the same stellar mass \citep{wetzel_etal12a}.

The similarity between satellite and central $M_\ast-M_h$ relations
thus appears to be a natural outcome of several facts: 1) the weakly
or unevolving $M_\ast-M_h$ relation for central galaxies; 2) the
hierarchical growth of structure, which ensures that the majority of
satellite galaxies (at least with $M^\ast>10^{10}\Msun$) are not
strongly affected by their environments for most of their relatively
short lives; 3) typical SFRs at late times are low, resulting in long
mass doubling times; and 4) the combination of the relatively short
satellite lifetimes and long mass doubling times suggests that the
typical satellite will not have grown substantially in mass since the time it became a satellite, and therefore the satellite relation will not differ
significantly from the central galaxy relation when the satellite was
accreted, which in turn differs little from the present day central
relation.


\subsection{Comparison to Previous Work}
\label{subsec:prev_work}


There have been several previous studies that attempted to separately
constrain the satellite and central galaxy $M_\ast-M_h$
relations. \cite{wang06} were the first to consider such a
distinction.  They adopted a method that used galaxy positions and
velocities determined from the orbital and merging histories of halos
and subhalos from a full cosmological simulation (as opposed to a
statistical HOD approach) to predict observed galaxy statistics.
Galaxy properties such as luminosity and stellar mass were
parameterized as a function of $\Macc$, where the functional forms of
the $L - M_h$ and $\Mstar - M_h$ relations were adopted from a
semi-analytic model as reference.  They were able to successfully
model the observed stellar mass/luminosity functions and 2PCFs at low
redshift and found central and satellite $\Mstar - M_h$ relations that
were identical within the errors.  These results are broadly
consistent with our work, although recent high-precision measurements
of $\wprp$ and $\fsat$ argue for a slight difference between the
central and satellite $M_\ast-M_h$ relations.

Both \citet{yang12} and \citet{moster13} treat 
satellite galaxies `self-consistently', wherein the $M_\ast-M_h$ relation for satellites is used at the epoch of accretion (for instance, see the multi-epoch abundance matching technique described in \citealt{moster13}).  Both of these studies find that the 
the global $M_\ast-M_h$ relation evolves with redshift (albeit 
weakly), thus it is also inferred that at the 
$z\sim0$ central and satellite relations are different. 

\citet{rod_puebla12} also considered distinct $M_\ast-M_h$ relations
for centrals and satellites in an abundance matching context.  They
used the $z\sim0$ GSMF decomposed into central and satellite galaxies
and used abundance matching separately for the centrals and
satellites.  They found models based on $\Macc$ were in agreement with
measured conditional stellar mass functions and correlation functions
only when the satellite $M_\ast-M_h$ relation was slightly different
from the central relation (by $\lesssim0.1$ dex at $M>10^{12}\Msun$),
a result in general accord with our low-$z$ findings.

\citet{neistein12} used abundance matching within a full cosmological
simulation to further investigate the $M_\ast-M_h$ connection.
Motivated by \citet{neistein11b}, satellite stellar mass depended on
both $\Macc$ and the host halo mass at $z=0$, and the satellite
$M_\ast-M_h$ relation was allowed to differ from the central relation.
The authors considered the abundance of galaxies, the 2PCF and weak
lensing measurements in their modeling and they found that the
$M_\ast-M_h$ relation for satellites at $z=0$ was poorly constrained,
such that a wide range of relations were consistent with the data,
resulting in satellite fractions spanning $\fsat = 0.05 - 0.8$.  This
result suggested that the inclusion of the measured satellite fraction
as an additional statistic could be key to further constraining the
satellite $M_\ast-M_h$ relation.

Each of these studies attempted to separately constrain the satellite
and central galaxy $M_\ast-M_h$ relations at low redshift, allowing us
to piece together a clearer picture of the two relations. While there
are several differences in the methodology between our work and the
aforementioned studies, the principal distinction is our use of the
measured satellite fractions from \citet{reddick12} as an additional
model constraint. The power of using satellite fractions to constrain
the satellite $M_\ast-M_h$ relation can also be seen in our Figure 1.
However, at $z\sim1$ satellite fractions are not available, and our
derived satellite $M_\ast-M_h$ relation is consequently more
uncertain.  In fact, the uncertainties we derive are qualitatively
consistent with the error envelope quoted by \citet{neistein12} based
on their analysis of $z\sim0$ data, also without satellite fractions.
Additionally, one can see from Figure 3 of \citet{wang_etal12} that
the successful modeling of 2PCFs from \citet{wang06} over-predicts
satellite fractions from the group catalogs of \citet{yang08}.  Our
results argue that the observed 2PCF and GSMF are not enough to
provide strong constraints on the satellite $M_\ast-M_h$ relation.
However, estimates of the mass-dependent satellite fraction, in
conjunction with these other statistics, result in much smaller errors
on the inferred satellite $M_\ast-M_h$ relation.

It is also important to emphasize here that we do not allow freedom in
the central relation in order to limit computation time.  We argue
that this approximation is justified because the global GSMF and
2PCFs on large scales (where central galaxies dominate) are well-fit
by our model.  However, incorporating a parameterized $M_\ast-M_h$
central relation may result in a larger range of possible models for
the satellites.  We leave this possibility for future work.

Finally, \citet{reddick12} showed that $\vpeak$, the maximum circular
velocity analog to $\Mpeak$, was the best halo property for matching
low-$z$ clustering data.  We have adopted the latter herein in order
to exploit the newly calibrated $M_\ast-M_h$ from B12, which was
derived for masses rather than circular velocities.  It remains an
open question if the satellite and central relations would be even
more similar if circular velocities were used in place of halo masses.


\section{SUMMARY}\label{sec:conclusion}


In this paper we have explored the relation between satellite galaxies
and their dark matter halos since $z\sim 2$.  We take advantage of
precision measurements of the galaxy 2PCF, a statistic which is highly
sensitive to the \emph{satellite} galaxy population.  Previous studies
have used low- and intermediate-redshift clustering measurements
\citep[see][]{moster10,yang12} as the 2PCF provides a powerful
constraint to better understand the evolution of stellar mass in
satellite galaxies.  For the first time we have incorporated
clustering measurements as a function of stellar mass out to high
redshift, allowing us to model the evolution of satellite galaxies over
$\sim10 \ \Gyr$ of cosmic time.     

To model the satellite galaxies, we started by adopting the
$M_\ast-M_h$ relation of B12 which was shown to accurately reproduce
GSMFs out to $z\sim 8$.  We applied this relation to halos and
subhalos in the Bolshoi dissipationless $N-$body simulation identified
by the ROCKSTAR halo finder at $z\sim 0,1$ and 2.  We then calculated
the projected 2PCFs to compare to observations at $z\sim 0$ and 1 as
well as angular correlation functions at $z\sim 2$.  This approach was
named the \emph{Sats=Cens} model as the B12 $M_\ast-M_h$ relation
gives the same functional form for both central and satellite
galaxies.  The \emph{Sats=Cens} model was shown to slightly
under-predict satellite fractions and clustering measurements in
low-mass galaxies at low redshift.  This prompted us to introduce a
model that allowed for more flexibility in the satellite $M_\ast-M_h$
relation, the \emph{varySats} model.  We retained the B12 best-fit
relation for centrals and fit for the satellite relation, using an
MCMC method to probe parameter space.  Our technique is essentially a
hybrid between the halo occupation distribution and abundance matching
methods, in the sense that we adopt the abundance matching requirement
that satellites reside within subhalos while allowing for a
flexibility in the galaxy-halo relation common to halo occupation
models.  Our modeling approach yields the following principal results.

\begin{itemize}

\item{The clustering and satellite fraction data at $z\sim0$ imply
    that the satellite galaxy $M_\ast-M_h$ relation is nearly
    identical to the central galaxy relation, with a preference for
    the stellar mass of satellites to be $\sim10$\% larger than
    centrals at fixed halo mass (the two relations are consistent at
    $2-3\sigma$ for subhalo masses $\gtrsim 10^{13}\Msun$).  At
    $z\sim1$ the satellite and central $M_\ast-M_h$ relations are
    indistinguishable within the (considerable) errors.}

\item{Our models provide an excellent fit to the stellar-mass
    dependent clustering of galaxies at $z\sim1$.  This stands in
    contrast to several previous studies that were unable to reproduce
    the observed $z\sim 1$ $B$-band luminosity-dependent clustering
    through abundance matching \citep{wetzel_white10,gerke12}.  The
    success herein may perhaps be due to the fact that stellar mass is
    more strongly correlated with dark matter halo mass than $B$-band
    luminosity.}

\item{At $z\sim 2$, we find that a simple model in which the satellite
    $M_\ast-M_h$ relation is the same as the central relation is in
    good agreement with two-point clustering data at all scales and
    all stellar masses.  However, at present the errors on the data
    are too large to constrain the more flexible satellite model.}

\end{itemize}

While the detailed physics of galaxy formation is complex and highly
nonlinear, our results add to the emerging consensus that the basic
properties of galaxies are surprisingly regular and simply connected
to the underlying dark matter distribution.  The $M_\ast-M_h$ relation
for central galaxies has been shown to evolve little since $z\sim 4$
\citep{yang12, moster13, behroozi12}. Along with this lack of
evolution, \citet{behroozi12b} showed that there exists a nearly
constant peak in star formation efficiency in halos with
characteristic mass $10^{11.7}$ since $z\sim 4$, consistent with
theoretical predictions of the halo mass scale for the shock heating
of accreted gas \citep{rees_ostriker77, whiterees78,blumenthal_etal84, keres05, dekel_birnboim06}.  Additionally, this is also the mass scale when effects due to AGN
feedback become significant, which can have  a dramatic effect on star
formation \citep{teyssier11,martizzi_etal12}. These results for
central galaxies also appear to hold for the satellites.  The reason
for this is not yet entirely clear, but it is likely related to the
fact that the typical $\sim L^\ast$ satellite galaxy resides in a poor
group where transformation processes are weak and lifetimes are short.


\acknowledgments

This material is based upon work supported by the National Science
Foundation under Award No. AST-1202698. We would like to thank the
referee for a careful and highly constructive referee report. We would
also like to thank Andreas Berlind, Andrew Wetzel, Martin White,
Rachel Reddick, and Risa Wechsler for insightful discussions and
Joanne Cohn, Peter Behroozi, Andrey Kravtsov and Andrew Hearin for
helpful comments on an earlier draft.  Finally, DFW and CC would also
like to thank Rachel Reddick, Nick Mostek and David Wake for providing
their data in electronic format and Peter Behroozi for access to the
Bolshoi halo catalogs.


\bibliography{./citations.bib}

\end{document}